\documentclass[aps,prx,twocolumn,%longbibliography,
superscriptaddress,floatfix,nofootinbib]{revtex4-2}

\usepackage{epsfig,amsmath,amssymb,color,comment,physics}
\usepackage[makeroom]{cancel}
\usepackage[caption=false]{subfig}
\usepackage{mathrsfs}
\usepackage[countmax]{subfloat}
\usepackage[normalem]{ulem}
\usepackage[english]{babel}
\usepackage{dsfont}
\usepackage[bookmarks=true,colorlinks,linkcolor=black,urlcolor=NavyBlue,citecolor=RoyalBlue]{hyperref}
\usepackage{mathtools}
\usepackage{siunitx}
\usepackage{upgreek}
\usepackage{comment}
\usepackage{xcolor}

\hypersetup{
    unicode=false,     % non-Latin characters in Acrobat bookmarks
    pdftoolbar=false,  % show Acrobat toolbar?
    pdfmenubar=true,   % show Acrobatmenu?
    pdffitwindow=false, % window fit to page when opened
    pdfstartview={FitH},% fits the width of the page to the window
    pdftitle={},    % title
    pdfauthor={Authors},     % author
    pdfsubject={},   % subject of the document
    pdfcreator={},   % creator of the document
    pdfproducer={}, % producer of the document
    pdfkeywords={quantum many-body scars} {superconducting processor} {quantum state tomography}, % list of keywords
    pdfnewwindow=true,% links in new window
    colorlinks=true,% false: boxed links; true: colored links
    linkcolor=black,% color of internal links (change box color with linkbordercolor)
    citecolor=blue, % color of links to bibliography
    filecolor=magenta,% color of file links
    urlcolor=blue% color of external links
}

\newcommand{\Er}{\textrm{E}_{\textrm{r}}}

\begin{document}

\title{Observing the emergence of a velocity hierarchy in matter waves}%strongly interacting bosons }

\author{ Xudong Yu } \thanks{These authors contributed equally to this work.}
\affiliation{Institut f{\"u}r Experimentalphysik und Zentrum f{\"u}r Quantenphysik, Universit{\"a}t Innsbruck, Technikerstra{\ss}e 25, Innsbruck, 6020, Austria} 

\author{ Wenhan Chen } \thanks{These authors contributed equally to this work.}
\affiliation{Institut f{\"u}r Experimentalphysik und Zentrum f{\"u}r Quantenphysik, Universit{\"a}t Innsbruck, Technikerstra{\ss}e 25, Innsbruck, 6020, Austria} 
\affiliation{Key Laboratory of Quantum State Construction and Manipulation (Ministry of Education), School of Physics, Renmin University of China, Beijing 100872, China}

\author{ Igor Zhuravlev }
\affiliation{Institut f{\"u}r Experimentalphysik und Zentrum f{\"u}r Quantenphysik, Universit{\"a}t Innsbruck, Technikerstra{\ss}e 25, Innsbruck, 6020, Austria} 

\author{ Yi Zeng }
\affiliation{Institut f{\"u}r Experimentalphysik und Zentrum f{\"u}r Quantenphysik, Universit{\"a}t Innsbruck, Technikerstra{\ss}e 25, Innsbruck, 6020, Austria} 

\author{ Sudipta Dhar}
\affiliation{Institut f{\"u}r Experimentalphysik und Zentrum f{\"u}r Quantenphysik, Universit{\"a}t Innsbruck, Technikerstra{\ss}e 25, Innsbruck, 6020, Austria}

\author{ Milena Horvath}
\affiliation{Institut f{\"u}r Experimentalphysik und Zentrum f{\"u}r Quantenphysik, Universit{\"a}t Innsbruck, Technikerstra{\ss}e 25, Innsbruck, 6020, Austria}

\author{ Thierry Giamarchi}
\affiliation{DQMP, University of Geneva, 24 Quai Ernest-Ansermet, Geneva, CH-1211, Switzerland}

\author{Laurent Sanchez-Palencia}
\affiliation{CPHT, CNRS, École Polytechnique, Institut Polytechnique de Paris, Palaiseau, France}

\author{ Manuele  Landini }
\affiliation{Institut f{\"u}r Experimentalphysik und Zentrum f{\"u}r Quantenphysik, Universit{\"a}t Innsbruck, Technikerstra{\ss}e 25, Innsbruck, 6020, Austria}

\author{ Hanns-Christoph  N{\"a}gerl}\email{christoph.naegerl@uibk.ac.at}
\affiliation{Institut f{\"u}r Experimentalphysik und Zentrum f{\"u}r Quantenphysik, Universit{\"a}t Innsbruck, Technikerstra{\ss}e 25, Innsbruck, 6020, Austria}

\author{ Yanliang  Guo }\email{yanliang.guo@uibk.ac.at}
\affiliation{Key Laboratory of Quantum State Construction and Manipulation (Ministry of Education), School of Physics, Renmin University of China, Beijing 100872, China}
\affiliation{Institut f{\"u}r Experimentalphysik und Zentrum f{\"u}r Quantenphysik, Universit{\"a}t Innsbruck, Technikerstra{\ss}e 25, Innsbruck, 6020, Austria}

\date{\today}

%\begin{document}

\begin{abstract}

Classical waves in dispersive media naturally exhibit distinct phase and group velocities%~\cite{brillouin1960}
. Whether an analogous separation of velocities can emerge in matter waves under strong many-body interactions has remained experimentally unexplored. Here, we demonstrate the emergence of a velocity hierarchy in a strongly interacting lattice gas. Using quench spectroscopy %~\cite{villa2019,menu2018,villa2020,chen2025,yu2025}
together with time-resolved correlation measurements, we independently determine the sound, group, and phase velocities across the superfluid–to-Mott-insulator transition. These velocities are nearly degenerate close to the transition, but progressively separate as the Mott gap opens and the quasiparticle dispersion acquires a massive relativistic-like form%~\cite{bloch2008many,Peter2012,cevolani2018,despres2019,schneider2021}
. Strikingly, phase-coherence fronts propagate faster than the Lieb–Robinson velocity scale %~\cite{Lieb1972}
while remaining fully consistent with locality. The measured velocities satisfy a relativistic-like invariance relation in the insulating regime. Our results establish propagation-velocity hierarchies as emergent signatures of strongly correlated quantum dynamics.

\end{abstract}
\maketitle
How information propagates in a medium is one of the defining questions of non-equilibrium many-body physics. In local quantum systems, causality constrains the spreading of disturbances through an effective light cone set by the Lieb-Robinson bound~\cite{Lieb1972}. Yet, observable dynamics need not be governed by a single propagation velocity. In dispersive media, the propagation of energy, phase, and correlations can separate dynamically, leading to distinct characteristic velocities for the propagation of sound, wave packets, and phase fronts~\cite{brillouin1960,anderson1958,pozar2011,Tonks1929}.  

%How \lsp{information propagates \sout{excitations and disturbances}} in a medium is one of the defining questions of non-equilibrium many-body physics. In local quantum systems, causality constrains the spreading of disturbances through an effective light cone set by the Lieb-Robinson bound~\cite{Lieb1972}. Yet observable dynamics need not {\color{blue}\sout{unfold through}be governed by} a single propagation {\color{blue}\sout{scale} velocity}. In dispersive media, the propagation of energy, phase, and correlations can separate dynamically, leading to distinct characteristic velocities for the propagation of sound, wave packets, and phase fronts~\cite{brillouin1960,anderson1958,pozar2011,Tonks1929}

\begin{figure*}[t!]
    \centering
    \includegraphics[width=1\linewidth]{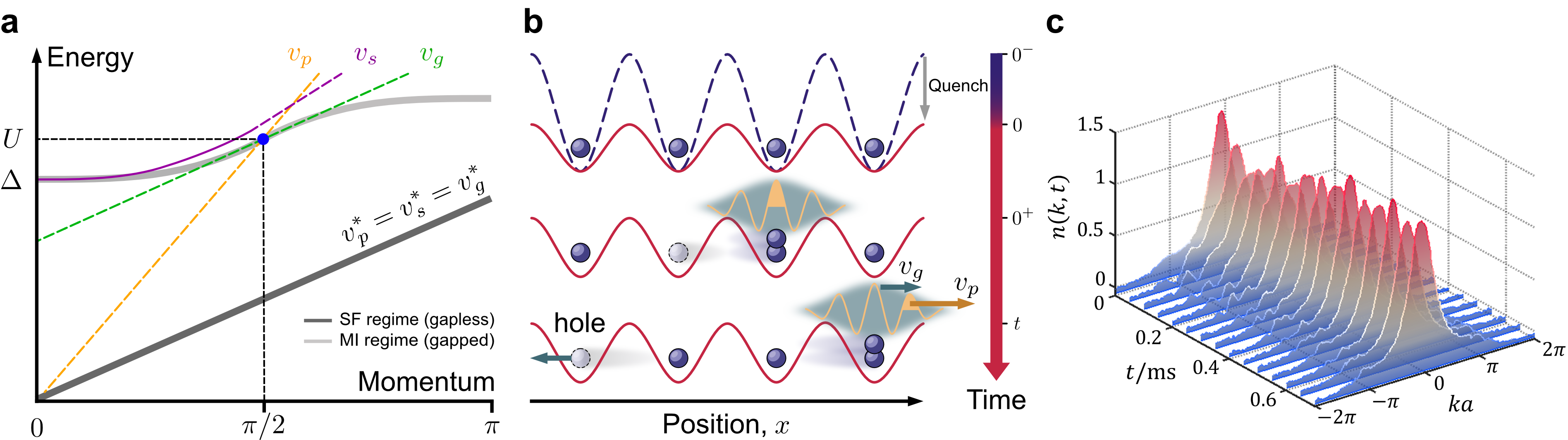}
    \caption{\textbf{Mechanism of velocity separation, sketch of the experimental protocol, and typical dataset.} (a) Not-to-scale schematic of the low-energy quasiparticle dispersion relation, illustrating the evolution from an almost linear dispersion in the SF regime (dark gray line) to a gapped, relativistic-like form in the MI regime (light gray curve). The Mott gap $\Delta$ and interaction energy $U$ are indicated on the vertical axis. The dashed lines represent the geometric definitions of the phase $v_p$, sound $v_s$, and group velocity $v_g$, where $v_s$ is defined in the MI dispersion $E(k)\!=\!\sqrt{\Delta^2+(\hbar v_sk)^2}$ and captured by the asymptote when $k\!\to\! \infty$. When the dispersion is linear, the three velocities are degenerate; as the gap opens and the dispersion bends, they separate into a hierarchy. (b) Illustration of the quench dynamics. The wave-packet motion of the doublon and holon after the lattice quench defines the group velocity, whereas the interference gives rise to propagating phase fronts with velocity $v_p$, which may exceed $v_g$. (c) Examples of measured post-quench momentum distributions $n(k,t)$ for $\gamma \approx 4.2$ in \SI{50}{\micro\second} time steps. }
    \label{fig:fig1}
\end{figure*}

While distinct propagation velocities are a hallmark of classical dispersive media, it is unknown whether strong many-body interactions can generate an analogous hierarchy in quantum matter. Strongly correlated lattice bosons described by the Bose–Hubbard (BH) model  provide an ideal setting to address this question, because interactions continuously reshape the low-energy quasiparticle dispersion from nearly linear to massive relativistic-like, naturally giving rise to distinct propagation velocities~\cite{bloch2008many,Peter2012,cevolani2018,despres2019,schneider2021}. As the interaction-induced gap opens, the quasiparticle dynamics progressively depart from the nearly linear regime near the transition, which should result in different values for the phase velocity $v_p$, sound velocity $v_s$ and group velocity $v_g$, see Fig.~\ref{fig:fig1}a and Supplementary Materials~\cite{supplement}. Previous experiments have revealed light-cone-like dynamics in quantum gases, ions, and spin systems~\cite{Cheneau2012,Jurcevic2014,richerme2014,Takasu2020,Fukuhara2013-gi}. However, these studies neither connected the observed propagation fronts to independently measured quasiparticle energy scales nor established how a velocity hierarchy emerges continuously as the many-body gap opens~\cite{Ali2021}. 

%\sout{However, these works did not directly connect the observed propagation fronts to independently measured quasiparticle dispersions, nor establish how a velocity hierarchy emerges continuously from the interaction-driven opening of a many-body gap}}. 

%\sout{Strongly correlated lattice bosons provide an ideal setting for the investigation into propagation speeds, as the tuning of interactions continuously reshapes the low-energy dispersion relation and thereby leads to the emergence of distinct propagation velocities }.}

%{\color{blue}\sout{In the Bose-Hubbard setting, increasing interactions drives the system from a gapless superfluid to a gapped Mott insulator, where the low-energy quasiparticle dispersion relation acquires a relativistic-like massive form} 

Here, we combine quench spectroscopy~\cite{villa2019,menu2018,villa2020,chen2025,yu2025} with time-resolved correlation measurements to reveal a velocity hierarchy and directly connect it to the underlying quasiparticle dispersion. Using interaction-tunable one-dimensional (1D) bosons, we reconstruct the full dispersion relation across the interaction-driven SF-MI transition and directly track the opening of the excitation gap. From the gap and the equilibrium correlation length we obtain $v_s$. Independently, the quench-induced space-time evolution of the one-body correlation function reveals a propagating envelope with phase-interference fringes inside, allowing us to read off $v_g$ and $v_p$. The three velocities form a clear hierarchy as interactions are increased and the gap widens. They are found to satisfy a relativistic-like invariance relation.

 The experiment starts by loading a Bose-Einstein condensate (BEC) of about $1.3\times10^5$ $^{133}$Cs atoms into an array of vertically oriented 1D tubes formed by a two-dimensional optical lattice in the horizontal $y$-$z$ plane. The lattice depth is ramped adiabatically to $25\,\Er$, where $\Er\!=\!\pi^2\hbar^2/(2ma^2)$ is the recoil energy and $a\!=\!\lambda/2\!=\!532.25\,\mathrm{nm}$ is the lattice spacing. Typically, about $5.5\times10^3$ tubes are populated, each containing on average $23$ atoms, with weak harmonic confinement along the longitudinal $x$-direction at frequency $\omega_x\!=\!2\pi\times 14.1(4)\,$Hz. A lattice along this axis is then ramped up to $V_x\!=\!20\,\Er$. Having set the scattering length to $a_s\!=\!300\,a_0$ via Feshbach-resonance tuning~\cite{Chin2010RMP}, we thereby prepare a 3D MI state with near unity filling. We next reduce $V_x$ adiabatically to 15 $\Er$, and then, as illustrated in Fig.~\ref{fig:fig1}b, quench the system to $5\,\Er$ within about \SI{200}{\micro\second}. After a variable evolution time $t$, we measure the post-quench momentum distribution $n(k,t)$, by switching off all optical fields and the inter-particle interaction and detecting the atoms via absorption imaging after $50\,$ms time of flight. An example of the time evolution of $n(k,t)$ is shown in Fig.~\ref{fig:fig1}c. Fourier transforming along $t$ yields the quench spectral function (QSF) $S(k,\nu)$ and hence the quasiparticle dispersion relation~\cite{villa2019}, while Fourier transforming along $k$ gives the time-dependent one-body correlation function $G^{(1)}(x,t)=\langle \psi^\dagger(x,t)\psi(0,t)\rangle$, with $\psi(x,t)$ the field operator at position $x$ and time $t$~\cite{bloch2008many}. For the measurement of $S(k,\nu)$ and $G^{(1)}(x,t)$, we typically take 60 equidistant time steps out to $10$ and $1.5$ ms, respectively.

%As shown in Fig.~\ref{fig:fig1}c, it encodes both correlation spreading and the excitation spectrum. In two complementary measurements detailed in the following based on the same quench protocol, we probe $n(k,t)$ over different time windows and resolutions. A long-time dataset, Fourier transformed along $t$, yields the quench spectral function (QSF) $S(k,\nu)$ and hence the quasiparticle dispersion~\cite{villa2019}. A separate short-time dataset, Fourier transformed along $k$, gives the real-space dynamics of the one-body correlations function $G^{(1)}(x,t)=\langle \psi^\dagger(x,t)\psi(0,t)\rangle$, with $\psi(x,t)$ the field operator at position $x$ and time $t$.

%This generates particle-hole pairs, which propagate throughout the system, generating a signal characterized by group and phase velocities.

%Fourier transformation along $k$ gives the real-space dynamics of the one-body correlations $G^{(1)}(x,t) = \langle \psi^\dagger(x,t) \psi(0,t)\rangle$, with $\psi(x,t)$ the field operator at position $x$ and time $t$,whereas Fourier transformation along $t$ yields the quench spectral function (QSF) $S(k, \nu)$ and hence the quasiparticle dispersion~\cite{villa2019}. 
\begin{figure*}[ht!]
    \centering
    \includegraphics[width=1\linewidth]{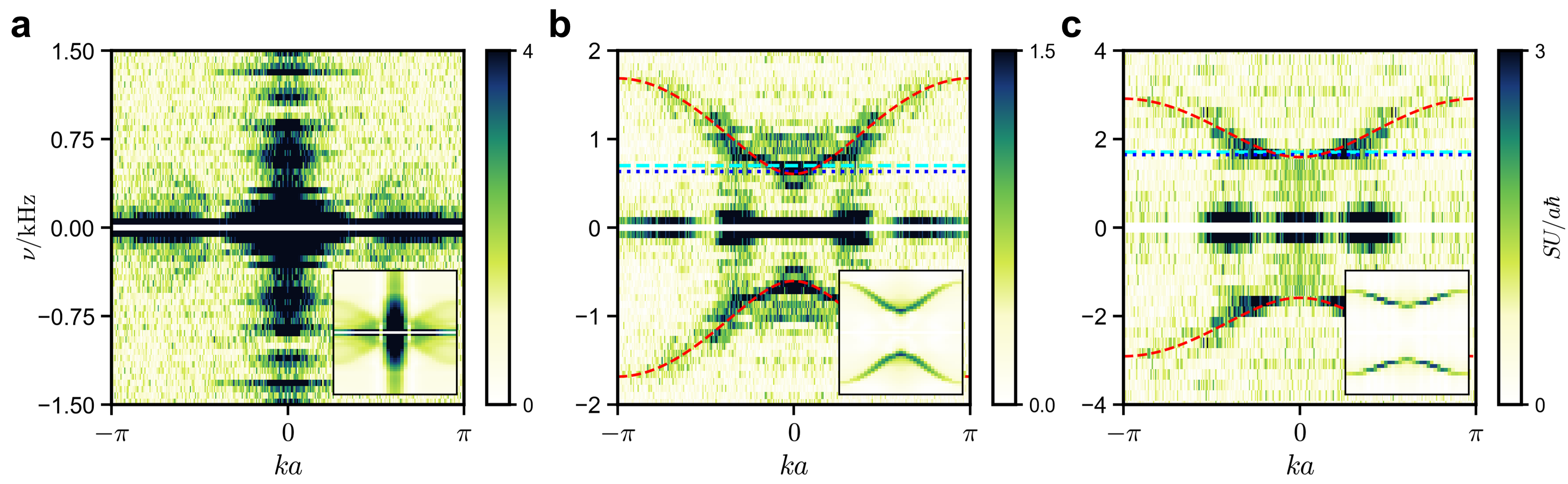}
    \caption{\textbf{Dispersion relations as obtained from quench spectroscopy.} Quench spectral function $S(k,\nu)$ for $\gamma \approx 0.3~{\rm (a)}$,$~4.2~{\rm (b)}$ and$~11.8~{\rm (c)}$ at $V_x=5\,\Er$, with $\gamma$ calculated assuming unit filling. The cyan dashed line and the blue dotted line in (b) and (c) indicate the energy gap extracted from LMS measurements and QMC calculations, respectively, where $\nu_{\rm LMS}\!=\!0.70(7)\,{\rm kHz}$ and $\nu_{\rm QMC}=0.63\,{\rm kHz}$ for (b), and $\nu_{\rm LMS}=1.70(14)\,{\rm kHz}$ and $\nu_{\rm QMC}=1.65\,{\rm kHz}$ for (c). The red dashed lines in (b) and (c) are fits to the measured dispersion relation. The insets show the results of the corresponding DMRG simulations.}
    \label{fig:fig2}
\end{figure*}

%Examples of $G^{(1)}(x,t)$ and $S(k, \nu)$ are shown in Figs.~\ref{fig:fig1} c and d, respectively.\\
\begin{figure*}[ht]
    \centering
    \hspace*{-0.5cm}
    \includegraphics[width=1\linewidth]{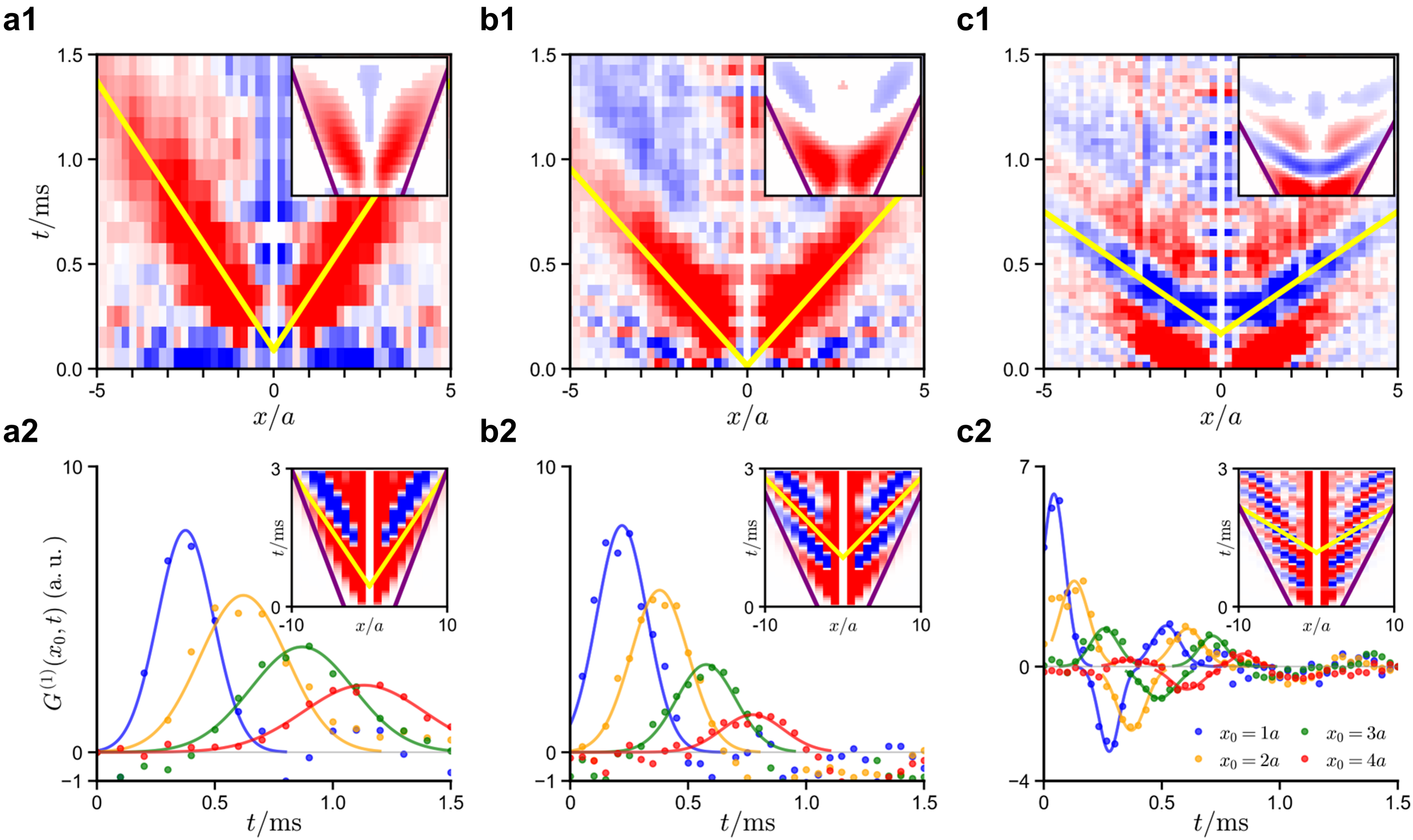}
    \caption{\textbf{Correlation-cone dynamics of $G^{(1)}(x,t)$.} (a1-c1) Space-time maps of $G^{(1)}(x,t)$ for $\gamma\approx2.5$ (a), $\gamma\approx4.2$ (b) and $\gamma\approx11.8$ (c). Yellow solid lines are linear fits to the trajectories of local extrema. Red and blue regions denote positive and negative values, respectively. Insets in (a1-c1) show the envelope structures extracted from the correlation-cone dynamics by convolution with a 2D Gaussian and application of a hard threshold. Purple solid lines are linear fits to the envelope propagation. (a2-c2) Temporal traces of $G^{(1)}(x,t)$ at selected positions $x_0=1a$ (blue), $2a$ (orange), $3a$ (green) and $4a$ (red). Solid lines are Gaussian fits. For (c2), the positive and negative peaks are fitted independently. Insets in (a2-c2) show the corresponding DMRG results using the values of $U$ and $J$ extracted independently from the dispersion relation measured under the same conditions. Yellow and purple solid lines indicate the phase and group velocities, respectively. The calculations are shown over twice the experimental space-time window, allowing a direct comparison of the propagation velocities. In c, the oscillation frequency averaged over different positions is $\nu=2.11(4)\,\mathrm{kHz}$, consistent with the independently extracted value $U/h=2.15(5)\,\mathrm{kHz}$ from the long-time dispersion measurement.}
    \label{fig:fig3}
\end{figure*}

We model the post-quench dynamics along the tube direction by the well-known 1D BH Hamiltonian~\cite{Jaksch1998}
\begin{align}
\label{eq.1}
    \hat{H}=-J\sum_i\left(\hat{a}_i^{\dagger}\hat{a}_{i+1}+\rm{H.c.}\right)+\frac{U}{2}\sum_i\hat{a}_i^{\dagger}\hat{a}_i^{\dagger}\hat{a}_i\hat{a}_i \nonumber
\\
+\sum_i [V_{\rm ext}(i)-\mu] \hat{a}_i^{\dagger}\hat{a}_i,
\end{align}
where $\hat a_i^\dagger$ and $\hat a_i$ are the bosonic creation and annihilation operators at lattice site $i$, $\mu$ is the chemical potential, and $V_{\rm ext}(i)$ accounts for the weak longitudinal confinement. For $V_x\!=\!5\,\Er$, the SF-MI transition is known to occur at a critical interaction strength $\gamma\!\approx\!0.66$, where $\gamma$ is the dimensionless 1D Lieb-Liniger interaction parameter~\cite{haller2010pinning,boeris2016}. By tuning $\gamma$ from $0.3$ to $21.3$ (via setting $a_s\!=\!30(2)\,a_0$ to $800(1)\,a_0$), we access both the SF and MI regimes of the post-quench system. For the time-dependent density-matrix renormalization-group (DMRG) calculations~\cite{white1992,schollwock2011}, the initial state is always taken as the ground state at $15\,\Er$, i.e., at the pre-quench lattice depth. The post-quench Hamiltonian is then chosen according to the interaction regime: in the deep MI regime, its renormalized single-band BH parameters $U$ and $J$ are extracted by fitting the measured dispersion relation with $E_{\rm ex}(k)\simeq U-6J\cos ka$~\cite{Peter2012}, see below, whereas in the SF regime we use the bare $s$-band BH parameters at $5\,\Er$.

%. {\color{blue}WC try: The residual horizontal features are tentatively associated with the finite-temperature characteristic frequency scale $f_T= 1/{(2\pi L_{\tau})}= k_B T/h\sim0.25\, {\rm kHz}$ for $T\!=\!15$nK arising from the finite imaginary-time direction of the underlying 1+1D conformal field theory~\cite{Sun2026}}.The residual horizontal features may reflect an additional low-frequency scale arising from finite temperature and finite system size~\cite{Sun2026,Cazalilla2011RMP}. For the experimental temperature $T\sim10\,\mathrm{nK}$, the corresponding thermal frequency $f_T \sim k_B T/h$ is about $0.2\,\mathrm{kHz}$, comparable to the observed spacing of these weak structures.

The measured $S(k,\nu)$ for selected values of the interaction strength is shown in Fig.~\ref{fig:fig2}. In the SF regime (Fig.~2a) we find significant spectral weight at low energies, consistent with the presence of gapless excitations. In contrast, in the MI regime (Figs.~2b,c) a dispersion-like feature appears in conjunction with a clear excitation gap. The gap increases with increasing interaction strength. Its value is in agreement with the results of lattice modulation spectroscopy (LMS)~\cite{haller2010pinning} and of quantum Monte Carlo (QMC) calculations. The results of the corresponding DMRG calculations, shown in the insets of Fig.~\ref{fig:fig2}, quantitatively reproduce the measured spectra on both sides of the transition. We note that the SF spectra feature some discrete structure. While this does not affect the extraction of the low-energy quasiparticle branch, its possible origin is discussed in the Supplementary Materials \cite{supplement}.

%The measured $S(k,\nu)$ for selected values of the interaction strength is shown in Fig.~\ref{fig:fig2}. In the SF regime \lsp{(Fig.~2a)} we find significant spectral weight at low energies, consistent with the presence of gapless excitations. In contrast, in the MI regime \lsp{(Figs.~2b,c)} a dispersion-like feature appears in conjunction with a clear excitation gap. The gap increases with increasing interaction strength. Its value is in agreement with the results of lattice modulation spectroscopy (LMS\lsp{, cyan dashed line})~\cite{haller2010pinning} and of quantum Monte Carlo (QMC\lsp{, blue dotted line}) calculations. The results of the corresponding DMRG calculations, shown in the insets of Fig.~\ref{fig:fig2}, quantitatively reproduce the measured spectra on both sides of the transition. We note that the SF spectra feature some discrete structure. While this does not affect the extraction of the low-energy quasiparticle branch, its possible origin is discussed in the supplemental material \cite{supplement}

In the SF regime, $v_s$ reduces to the sound velocity, whereas in the insulating regime it corresponds to the velocity parameter in the relativistic-like low-energy dispersion relation $E(k)=\sqrt{\Delta^2+(\hbar v_s k)^2}$.
We determine the characteristic velocity $v_s$ from the spectra through $1/\xi \sim \Delta/(\hbar v_s)$~\cite{Cazalilla2011RMP} obtained from the excitation gap $\Delta$ and the equilibrium correlation length $\xi$. Experimentally, we obtain $\xi$ from the exponential decay of the equilibrium one-body correlation function, $G^{(1)}(x)\sim e^{-x/\xi}$~\cite{Guo2025MBDL,guo2024anomalous}, extracted from the static $n(k)$ at $V_x=5\,\Er$ and a given $\gamma$. Combining $\xi$ with the independently measured gap yields $v_s$ for varying $\gamma$.

\begin{figure*}
    \centering
    \includegraphics[width=0.6\linewidth]{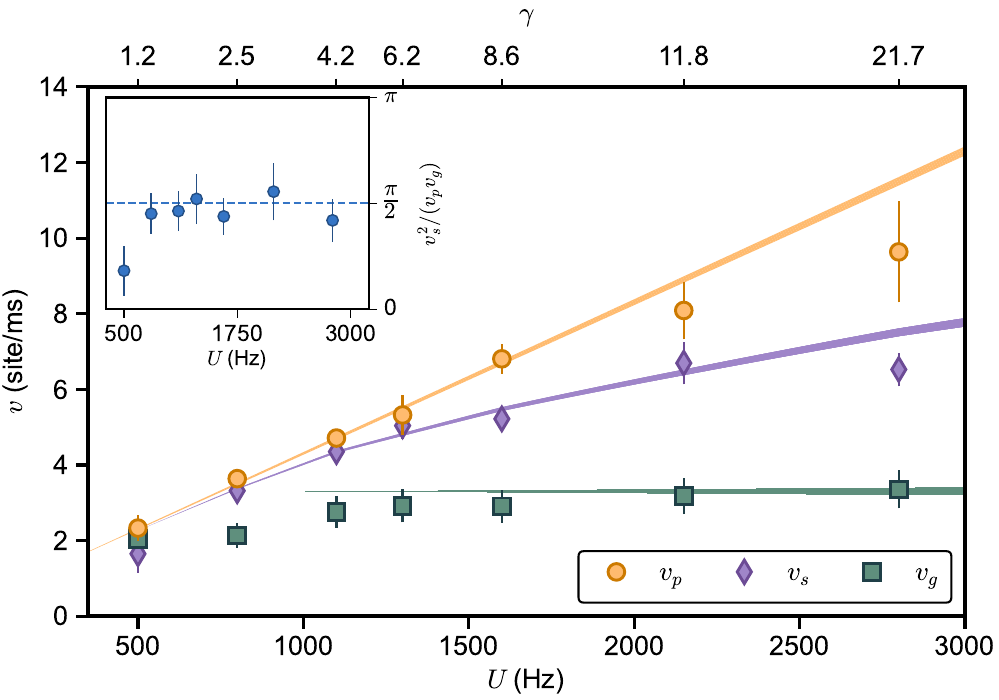}
    \caption{
    \textbf{Emergence of velocity hierarchy and relativistic-like invariance.} Comparison of the three characteristic velocities $v_p$ (circles), $v_s$ (diamonds) and $v_g$ (squares) as a function of $U$, with the corresponding values of $\gamma$ shown on the top axis. Their uncertainties are mainly from the fitting. The orange solid line indicates the linear relation $v_p\!=\!2U/\pi\hbar$ and the purple solid line are the calculated $v_s\!=\!\sqrt{6JU}/\hbar$. The green solid line denotes $v_g\!=\!6J/\hbar$ for the limit $U\!\rightarrow\!\infty$. Increasing line thickness indicates the primary regime of validity for each asymptotic expression. %The variation in line thickness indicates the expected validity range of the asymptotic expressions, with the thickest segments marking the regime where they are applicable. 
    Inset: interaction dependence of the ratio $v_s^2/(v_pv_g)$, obtained from the three independently measured velocities in this figure. The horizontal dashed line  marks the predicted $\pi/2$ value, indicating the relativistic-like invariance. 
    }
    \label{fig:fig4}
\end{figure*}
%Inset: two dispersion relations extracted from the measured QSF, for $U\sim1.1\,\mathrm{kHz}$ for $\gamma=4.2$ (light gray) and $U\sim2.2\,\mathrm{kHz}$ for $\gamma=11.8$ (dark gray). Yellow dashed lines indicate the corresponding $v_p$, and green dashed lines the corresponding $v_p$.

We next determine the phase and group velocities from the post-quench evolution of the one-body correlation function $G^{(1)}(x,t)$. The oscillatory interference fringes inside the correlation cone define the phase velocity $v_p$, whereas the envelope determines the group velocity $v_g$. A pronounced correlation cone emerges in the space-time plane, see Figs.~\ref{fig:fig3}a1-c1. As the interaction increases, the V-shaped Pattern of extrema progressively opens. The correlation pattern shows structures characterized by alternating positive and negative values. We quantify the speed of propagation by tracking the extrema of $G^{(1)}(x,t)$ within the cone. The trajectories are linear in distance and time, indicating  ballistic propagation, and their fitted slope defines $v_p$. Representative line cuts $G^{(1)}(x_0,t)$ at different lattice sites $x_0$ are shown in Figs.~\ref{fig:fig3}a2-c2. As the interaction strength increases, the propagation time over a fixed distance shortens. In addition, for the strongest interaction, $\gamma\approx11.8$, secondary oscillation fronts become resolved, with a frequency consistent with the interaction scale $U$ extracted from the independently measured dispersion relation under the same conditions. Over the full experimental range $\gamma\approx1.2$ to $21.7$, the extracted phase velocity increases systematically with interaction strength.

To extract the group velocity $v_g$ from the same measurement, we analyze the envelope of the post-quench correlation spreading. We first convolute $G^{(1)}(x,t)$ with a 2D Gaussian and then apply a hard threshold $\varepsilon$ to suppress weak residual signals (see Methods) \cite{supplement}. The resulting envelope dynamics are shown in the insets of Fig.~\ref{fig:fig3}a1-c1. The outer boundary of the envelope defines the maximal propagation speed in the lattice system, which we identify as $v_g$ and associate with the Lieb-Robinson velocity scale. Using this procedure, we determine $v_g$ over the full interaction range. These results are mirrored by our DMRG simulations using the values of $U$ and $J$ extracted from the measured $S(k,\nu)$. As shown in the insets of Fig.~\ref{fig:fig3}a2-c2, the numerics reproduce both the inner phase-front structure and the outer envelope of the measured correlation spreading. The corresponding $v_p$ and $v_g$ quantitatively agree with the experiment within $5\%$%The corresponding $v_p$ and $v_g$ are in quantitative agreement with the experiment within $5\%$%\lsp{~[with what precision?]}
, showing that increasing interactions progressively separate these characteristic propagation velocities and thereby establish a velocity hierarchy.

This independent determination allows us to directly compare the three propagation velocities without adjustable parameters and to establish their full interaction dependence. The data is summarized in Fig.~\ref{fig:fig4}. The three velocities evolve in qualitatively different ways as the interaction strength increases: $v_p$ increases most rapidly, $v_s$ also rises but more gradually, whereas $v_g$ shows only a weak initial increase before saturating. These three velocities characterize distinct complementary dynamical processes: $v_g$ governs the causality cone, i.e., the maximal spreading of the correlation envelope; $v_s$  defines the emergent low-energy quasiparticle scale; and $v_p$ governs the propagation of coherent interference fronts. As a result, the three velocities become progressively more separated with increasing interaction strength, leading to a clear velocity hierarchy $v_p>v_s>v_g$ in the deep Mott regime. Importantly, this hierarchy emerges continuously from a regime where all three velocities are nearly degenerate. This is quantitatively captured by the low-energy quasiparticle description. In the regime $U\!\gg \!J$, the analytical expressions for the three velocities reduce to $v_p\!=\!2U/\pi$, $v_s\!=\!\sqrt{6UJ}$, $v_g\!=\!6J$ (see Supplementary Materials). The independently measured three velocities agree well with these expressions, see Fig.\ref{fig:fig4}. The same description also implies a relativistic-like invariant relation for the three velocities, $v_s^2\!=\!(\pi/2)v_pv_g$, analogous to $c^2\!=\!v_pv_g$ for a massive relativistic particle, with the factor $\pi/2$ originating from the lattice periodicity. This relation is not imposed by the analysis but follows from the independently measured velocities. We plot $v_s^2/(v_pv_g)$ for varying interaction strength in the inset of Fig.~\ref{fig:fig4}. Once the interaction is strong enough, the ratio approaches the constant value $\pi/2$. This agreement provides direct experimental evidence for the validity of the relativistic-like low-energy description in the deep Mott regime.

In conclusion, we have experimentally shown that strong interactions reorganize quantum dynamics into a hierarchy of propagation velocities. Although causality remains governed by a single Lieb–Robinson velocity scale, phase, group and sound velocities separate continuously as the Mott gap develops while remaining quantitatively linked through a common quasiparticle description. These results establish propagation velocities as experimentally accessible probes of the internal dynamical organization of correlated quantum matter. Most strikingly, the phase-coherence front propagates faster than the Lieb-Robinson velocity scale while remaining fully consistent with locality. This approach opens a route to exploring coherence, transport, and information propagation in quantum many-body systems with multiple propagation channels, and can be extended to higher dimensions, to multiple excitation branches, and to settings where topology, frustration, disorder or long-range couplings generate even richer propagation structures~\cite{carleo2014,cevolani2018,frerot2018,villa2021b,villa2021a}.

\noindent{\bf Acknowledgments}\\
We acknowledge initial insightful discussions with H. Yao and T. Shi and support for the QMC calculation from C. Wu. The Innsbruck team acknowledges funding by the European Research Council (ERC) under project number 101201611 and by an FFG infrastructure grant with project number FO999896041. Y.G. is supported by the Austrian Science Fund (FWF) with project number 10.55776/COE1, and Quantum Science and Technology Major Project of China (Grant No.2025ZD0300400). L.S.P acknowledges funding from the Agence Nationale de la Recherche under projects QuanTEdu-France (No. ANR-CMAQ-002 France 2030) and QUTISYM (No. ANR-23-PETQ-0002), and GENCI-TGCC (Grant No. A0200510300). T.G. acknowledges funding from the Swiss National Science Foundation under grant number 200020-219400. Y.Z. is supported by the Austrian Academy of Sciences (\"OAW) with APART-MINT 12234.
\textbf{Author Contributions:} This work was conceived by Y.G., L.S.P., H.C.N. and T.G. Experiments were prepared by X.Y., I.Z., S.D., M.H. Y.Z. and performed by X.Y., I.Z. and Y.G. Data
were analyzed by W.C., X.Y., I.Z and Y.G. Numerical simulations and analytical predictions were performed by W.C. and L.S.P. The manuscript was drafted mainly by Y.G.,  W.C., X.Y. and H.C.N. All authors contributed to the discussion and finalization of the manuscript.
{\bf Data Availability:} The data shown in the main text are available via Zenodo~\cite{zenodo_data_availability}.
{\bf Code Availability:} The Density Matrix Renormalization Group and Time Evolving Block Decimation algorithm are based on the TeNPy project~\cite{tenpy2024}. Codes supporting the findings of this study are available from the corresponding author on a reasonable request.
\bibliography{cite}

@article{Ali2021,
  title = {Phase and group velocities for correlation spreading in the Mott phase of the Bose-Hubbard model in dimensions greater than one},
  author = {Mokhtari-Jazi, Ali and Fitzpatrick, Matthew R. C. and Kennett, Malcolm P.},
  journal = {Phys. Rev. A},
  volume = {103},
  issue = {2},
  pages = {023334},
  numpages = {10},
  year = {2021},
  month = {Feb},
  publisher = {American Physical Society}
}

@misc{zenodo_data_availability,
  note = {Data set is available from Zenodo at 10.5281/zenodo.21609850}
}

@article{yu2025,
  title = {Quench spectroscopy for {L}ieb-{L}iniger bosons in the presence of harmonic trap},
  author = {Yu, Jiachen and Hu, Yuanzhe and Chen, Wenhan and Yang, Jianing and Chen, Xuzong and Yao, Hepeng},
  journal = {Phys. Rev. Res.},
  volume = {7},
  issue = {4},
  pages = {L042066},
  numpages = {7},
  year = {2025},
  month = {Dec},
  publisher = {American Physical Society},
}

@book{brillouin1960,
  title={Wave Propagation and Group Velocity},
  author={Brillouin, L{\'{e}}on},
  year={1960},
  publisher={Academic Press},
  address={New York},
  volume={8},
  series={Pure and Applied Physics}
}

@book{pozar2011,
  title={Microwave Engineering},
  author={Pozar, David M.},
  year={2011},
  publisher={Wiley},
  edition={4th},
  address={Hoboken, NJ}
}

@article{Tonks1929,
  title = {Oscillations in Ionized Gases},
  author = {Tonks, Lewi and Langmuir, Irving},
  journal = {Phys. Rev.},
  volume = {33},
  issue = {2},
  pages = {195--210},
  numpages = {0},
  year = {1929},
  month = {Feb},
  publisher = {American Physical Society},
}

@article{anderson1958,
  title = {Coherent Excited States in the Theory of Superconductivity: Gauge Invariance and the {M}eissner Effect},
  author = {Anderson, P. W.},
  journal = {Phys. Rev.},
  volume = {110},
  issue = {4},
  pages = {827--835},
  numpages = {0},
  year = {1958},
  month = {May},
  publisher = {American Physical Society},
}

@article{Cazalilla2011RMP,
  title = {One dimensional bosons: From condensed matter systems to ultracold gases},
  author = {Cazalilla, M. A. and Citro, R. and Giamarchi, T. and Orignac, E. and Rigol, M.},
  journal = {Rev. Mod. Phys.},
  volume = {83},
  issue = {4},
  pages = {1405--1466},
  numpages = {0},
  year = {2011},
  month = {Dec},
  publisher = {American Physical Society}
}

@article{
chen2025,
author = {Cheng Chen  and Gabriel Emperauger  and Guillaume Bornet  and Filippo Caleca  and Bastien Gély  and Marcus Bintz  and Shubhayu Chatterjee  and Vincent Liu  and Daniel Barredo  and Norman Y. Yao  and Thierry Lahaye  and Fabio Mezzacapo  and Tommaso Roscilde  and Antoine Browaeys },
title = {Spectroscopy of elementary excitations from quench dynamics in a dipolar {X}{Y} {R}ydberg simulator},
journal = {Science},
volume = {389},
number = {6759},
pages = {483-487},
year = {2025}}

@article{despres2019,
  title={Twofold correlation spreading in a strongly correlated lattice {B}ose gas},
  author={Despres, Julien and Villa, Louis and Sanchez-Palencia, Laurent},
  journal={Scientific reports},
  volume={9},
  number={1},
  pages={4135},
  year={2019},
  publisher={Nature Publishing Group UK London}
}

@article{villa2019,
  title = {Unraveling the excitation spectrum of many-body systems from quantum quenches},
  author = {Villa, Louis and Despres, Julien and Sanchez-Palencia, Laurent},
  journal = {Phys. Rev. A},
  volume = {100},
  issue = {6},
  pages = {063632},
  numpages = {11},
  year = {2019},
  month = {Dec},
  publisher = {American Physical Society}
}

@article{villa2020,
  title = {Local quench spectroscopy of many-body quantum systems},
  author = {Villa, L. and Despres, J. and Thomson, S. J. and Sanchez-Palencia, L.},
  journal = {Phys. Rev. A},
  volume = {102},
  issue = {3},
  pages = {033337},
  numpages = {12},
  year = {2020},
  month = {Sep},
  publisher = {American Physical Society}
}

@article{menu2018,
  title = {Quench dynamics of quantum spin models with flat bands of excitations},
  author = {Menu, Rapha\"el and Roscilde, Tommaso},
  journal = {\Jprb},
  volume = {98},
  pages = {205145},
  numpages = {18},
  year = {2018}
}

@article{villa2021a,
  title = {Finding the phase diagram of strongly correlated disordered bosons using quantum quenches},
  author = {Villa, L. and Thomson, S. J. and Sanchez-Palencia, L.},
  journal = {Phys. Rev. A},
  volume = {104},
  issue = {2},
  pages = {023323},
  numpages = {17},
  year = {2021},
  month = {Aug},
  publisher = {American Physical Society}
}

@article{villa2021b,
  title = {Quench spectroscopy of a disordered quantum system},
  author = {Villa, L. and Thomson, S. J. and Sanchez-Palencia, L.},
  journal = {Phys. Rev. A},
  volume = {104},
  issue = {2},
  pages = {L021301},
  numpages = {6},
  year = {2021},
  month = {Aug},
  publisher = {American Physical Society}
}

@phdthesis{villa_thesis,
  TITLE = {{Out-of-equilibrium dynamics and quench spectroscopy of ultracold many-body quantum systems}},
  AUTHOR = {Villa, Louis},
  NUMBER = {2021IPPAX041},
  SCHOOL = {Institut Polytechnique de Paris},
  YEAR = {2021},
  MONTH = Jul,
  TYPE = {Theses},
  HAL_ID = {tel-03372978},
  HAL_VERSION = {v1}
}

@article{cevolani2018,
  title = {Universal scaling laws for correlation spreading in quantum systems with short- and long-range interactions},
  author = {Cevolani, Lorenzo and Despres, Julien and Carleo, Giuseppe and Tagliacozzo, Luca and Sanchez-Palencia, Laurent},
  journal = {Phys. Rev. B},
  volume = {98},
  issue = {2},
  pages = {024302},
  numpages = {9},
  year = {2018},
  month = {Jul},
  publisher = {American Physical Society}
}

@article{Cheneau2012,
  title={Light-cone-like spreading of correlations in a quantum many-body system},
  author={Cheneau, Marc and Barmettler, Peter and Poletti, Dario and Endres, Manuel and Schau{\ss}, Peter and Fukuhara, Takeshi and Gross, Christian and Bloch, Immanuel and Kollath, Corinna and Kuhr, Stefan},
  journal={\Jnature},
  volume={481},
  number={7382},
  pages={484--487},
  year={2012},
  publisher={Nature Publishing Group UK London}
}

@article{Peter2012,
  title = {Propagation front of correlations in an interacting Bose gas},
  author = {Barmettler, Peter and Poletti, Dario and Cheneau, Marc and Kollath, Corinna},
  journal = {Phys. Rev. A},
  volume = {85},
  issue = {5},
  pages = {053625},
  numpages = {14},
  year = {2012},
  month = {May},
  publisher = {American Physical Society}
}

@article{olshanii1998,
  title={Atomic scattering in the presence of an external confinement and a gas of impenetrable bosons},
  author={Olshanii, Maxim},
  journal={\Jprl},
  volume={81},
  number={5},
  pages={938},
  year={1998},
  publisher={APS}
}

@article{bloch2008many,
  title={Many-body physics with ultracold gases},
  author={Bloch, Immanuel and Dalibard, Jean and Zwerger, Wilhelm},
  journal={\Jrmp},
  volume={80},
  number={3},
  pages={885--964},
  year={2008},
  publisher={APS}
}

@article{guo2024anomalous,
  title={Anomalous cooling of bosons by dimensional reduction},
  author={Guo, Yanliang and Yao, Hepeng and Dhar, Sudipta and Pizzino, Lorenzo and Horvath, Milena and Giamarchi, Thierry and Landini, Manuele and N{\"a}gerl, Hanns-Christoph},
  journal={Science Advances},
  volume={10},
  number={7},
  pages={eadk6870},
  year={2024},
  publisher={American Association for the Advancement of Science}
}

@article{haller2010pinning,
  title={Pinning quantum phase transition for a {L}uttinger liquid of strongly interacting bosons},
  author={Haller, Elmar and Hart, Russell and Mark, Manfred J and Danzl, Johann G and Reichs{\"o}llner, Lukas and Gustavsson, Mattias and Dalmonte, Marcello and Pupillo, Guido and N{\"a}gerl, Hanns-Christoph},
  journal={\Jnature},
  volume={466},
  number={7306},
  pages={597--600},
  year={2010},
  publisher={Nature Publishing Group UK London}
}

@article{white1992,
  title = {Density matrix formulation for quantum renormalization groups},
  author = {White, Steven R.},
  journal = {Phys. Rev. Lett.},
  volume = {69},
  issue = {19},
  pages = {2863--2866},
  year = {1992}
}

@article{schollwock2011,
  title = {The density-matrix renormalization group in the age of matrix product states},
  journal = {Annals of Physics},
  volume = {326},
  pages = {96--192},
  year = {2011},
  author = {Ulrich Schollw\"ock}
}

@article{Jaksch1998,
  title = {Cold Bosonic Atoms in Optical Lattices},
  author = {Jaksch, D. and Bruder, C. and Cirac, J. I. and Gardiner, C. W. and Zoller, P.},
  journal = {Phys. Rev. Lett.},
  volume = {81},
  issue = {15},
  pages = {3108--3111},
  numpages = {0},
  year = {1998},
  month = {Oct},
  publisher = {American Physical Society}
}

@article{Chin2010RMP,
  title = {Feshbach resonances in ultracold gases},
  author = {Chin, Cheng and Grimm, Rudolf and Julienne, Paul and Tiesinga, Eite},
  journal = {Rev. Mod. Phys.},
  volume = {82},
  issue = {2},
  pages = {1225--1286},
  numpages = {0},
  year = {2010},
  month = {Apr},
  publisher = {American Physical Society}
}

@ARTICLE{Jurcevic2014,
  title     = "Quasiparticle engineering and entanglement propagation in a
               quantum many-body system",
  author    = "Jurcevic, P and Lanyon, B P and Hauke, P and Hempel, C and
               Zoller, P and Blatt, R and Roos, C F",
  journal   = "Nature",
  publisher = "Springer Science and Business Media LLC",
  volume    =  511,
  number    =  7508,
  pages     = "202--205",
  month     =  jul,
  year      =  2014,
}

@article{Cevolani2015,
  title = {Protected quasilocality in quantum systems with long-range interactions},
  author = {Cevolani, Lorenzo and Carleo, Giuseppe and Sanchez-Palencia, Laurent},
  journal = {Phys. Rev. A},
  volume = {92},
  issue = {4},
  pages = {041603},
  numpages = {5},
  year = {2015},
  month = {Oct},
  publisher = {American Physical Society}
}

@ARTICLE{Cevolani2016,
  title     = "Spreading of correlations in exactly solvable quantum models
               with long-range interactions in arbitrary dimensions",
  author    = "Cevolani, Lorenzo and Carleo, Giuseppe and Sanchez-Palencia,
               Laurent",
  journal   = "New J. Phys.",
  publisher = "IOP Publishing",
  volume    =  18,
  number    =  9,
  pages     = "093002",
  month     =  sep,
  year      =  2016,
  copyright = "http://creativecommons.org/licenses/by/3.0/"
}

@ARTICLE{Fukuhara2013-gi,
  title     = "Microscopic observation of magnon bound states and their
               dynamics",
  author    = "Fukuhara, Takeshi and Schau{\ss}, Peter and Endres, Manuel and
               Hild, Sebastian and Cheneau, Marc and Bloch, Immanuel and Gross,
               Christian",
  journal   = "Nature",
  publisher = "Springer Science and Business Media LLC",
  volume    =  502,
  number    =  7469,
  pages     = "76--79",
  month     =  oct,
  year      =  2013,
}

@Article{tenpy2024,
    title={{Tensor network Python (TeNPy) version 1}},
    author={Johannes Hauschild and Jakob Unfried and Sajant Anand and Bartholomew Andrews and Marcus Bintz and Umberto Borla and Stefan Divic and Markus Drescher and Jan Geiger and Martin Hefel and K\'evin H\'emery and Wilhelm Kadow and Jack Kemp and Nico Kirchner and Vincent S. Liu and Gunnar M\"oller and Daniel Parker and Michael Rader and Anton Romen and Samuel Scalet and Leon Schoonderwoerd and Maximilian Schulz and Tomohiro Soejima and Philipp Thoma and Yantao Wu and Philip Zechmann and Ludwig Zweng and Roger S. K. Mong and Michael P. Zaletel and Frank Pollmann},
    journal="SciPost Phys. Codebases",
    pages={41},
    year={2024},
    publisher={SciPost}
}

@article{
Guo2025MBDL,
author = {Yanliang Guo  and Sudipta Dhar  and Ang Yang  and Zekai Chen  and Hepeng Yao  and Milena Horvath  and Lei Ying  and Manuele Landini  and Hanns-Christoph Nägerl },
title = {Observation of many-body dynamical localization},
journal = {Science},
volume = {389},
number = {6761},
pages = {716-719},
year = {2025}}

@article{
Takasu2020,
author = {Yosuke Takasu  and Tomoya Yagami  and Hiroto Asaka  and Yoshiaki Fukushima  and Kazuma Nagao  and Shimpei Goto  and Ippei Danshita  and Yoshiro Takahashi },
title = {Energy redistribution and spatiotemporal evolution of correlations after a sudden quench of the {B}ose-{H}ubbard model},
journal = {Science Advances},
volume = {6},
number = {40},
pages = {eaba9255},
year = {2020}}

@ARTICLE{Lieb1972,
  title     = "The finite group velocity of quantum spin systems",
  author    = "Lieb, Elliott H and Robinson, Derek W",
  journal   = "Commun. Math. Phys.",
  publisher = "Springer Science and Business Media LLC",
  volume    =  28,
  number    =  3,
  pages     = "251--257",
  month     =  sep,
  year      =  1972
}

@misc{supplement,
    note={See {S}upplementary {M}aterials for details.}

}

@article{schneider2021,
  title = {Spreading of correlations and entanglement in the long-range transverse {I}sing chain},
  author = {Schneider, J. T. and Despres, J. and Thomson, S. J. and Tagliacozzo, L. and Sanchez-Palencia, L.},
  journal = {Phys. Rev. Res.},
  volume = {3},
  issue = {1},
  pages = {L012022},
  numpages = {7},
  year = {2021},
  month = {Mar},
  publisher = {American Physical Society}
}

@article{richerme2014,
  title={Non-local propagation of correlations in quantum systems with long-range interactions},
  volume={511},
  journal={\Jnature},
  author={Philip Richerme and Zhe-Xuan Gong and Aaron Lee and Crystal Senko and Jacob Smith and Michael Foss-Feig and Spyridon Michalakis and Alexey V. Gorshkov and Christopher Monroe}, 
  year={2014},
  pages={198--201}
}

@article{frerot2018,
  title = {Multispeed Prethermalization in Quantum Spin Models with Power-Law Decaying Interactions},
  author = {Fr\'erot, Ir\'en\'ee and Naldesi, Piero and Roscilde, Tommaso},
  journal = {\Jprl},
  volume = {120},
  pages = {050401},
  numpages = {6},
  year = {2018}
}

@article{carleo2014,
  title = {Light-cone effect and supersonic correlations in one- and two-dimensional bosonic superfluids},
  author = {Carleo, Giuseppe and Becca, Federico and Sanchez-Palencia, Laurent and Sorella, Sandro and Fabrizio, Michele},
  journal = {\Jpra},
  volume = {89},
  issue = {3},
  pages = {031602(R)},
  numpages = {5},
  year = {2014},
  publisher = {American Physical Society},
}

@article{boeris2016,
  title = {{M}ott transition for strongly interacting one-dimensional bosons in a shallow periodic potential},
  author = {Bo\'eris, G. and Gori, L. and Hoogerland, M. D. and Kumar, A. and Lucioni, E. and Tanzi, L. and Inguscio, M. and Giamarchi, T. and D'Errico, C. and Carleo, G. and Modugno, G. and Sanchez-Palencia, L.},
  journal = {\Jpra},
  volume = {93},
  issue = {1},
  pages = {011601(R)},
  numpages = {5},
  year = {2016}
}

@article{yan2015incorporating,
  title={Incorporating exact two-body propagators for zero-range interactions into N-body Monte Carlo simulations},
  author={Yan, Yangqian and Blume, D},
  journal={\Jpra},
  volume={91},
  number={4},
  pages={043607},
  year={2015},
  publisher={APS}
}

@inproceedings{troyer1998parallel,
  title={Parallel object oriented {M}onte {C}arlo simulations},
  author={Troyer, Matthias and Ammon, Beat and Heeb, Elmar},
  booktitle={International symposium on computing in object-oriented parallel environments},
  pages={191--198},
  year={1998},
  organization={Springer}
}

@article{albuquerque2007alps,
  title={The {ALPS} project release 1.3: Open-source software for strongly correlated systems},
  author={Albuquerque, A Fabricio and Alet, Fabien and Corboz, Philippe and Dayal, Prakash and Feiguin, Adrian and Fuchs, Sebastian and Gamper, L and Gull, Emanuel and G{\"u}rtler, S and Honecker, Andreas and others},
  journal={Journal of Magnetism and Magnetic Materials},
  volume={310},
  number={2},
  pages={1187--1193},
  year={2007},
  publisher={Elsevier}
}

@article{bauer2011alps,
  title={The {ALPS} project release 2.0: open source software for strongly correlated systems},
  author={Bauer, Bela and Carr, LD and Evertz, Hans Gerd and Feiguin, Adrian and Freire, J and Fuchs, Sebastian and Gamper, Lukas and Gukelberger, Jan and Gull, Emanuel and Guertler, Siegfried and others},
  journal={Journal of Statistical Mechanics: Theory and Experiment},
  volume={2011},
  number={05},
  pages={P05001},
  year={2011}
}

\cleardoublepage
\setcounter{figure}{0}         
\renewcommand{\thefigure}{S.\arabic{figure}} 

\section*{Supplementary materials}\label{sec:Methods}

\subsection{Experimental parameters}

Our experimental sequence begins with the production of a Bose-Einstein condensate (BEC) in a 3D optical dipole trap, with harmonic frequencies of $(\omega_x, \omega_y, \omega_z) \!=\! 2\pi \times (11.8(4), 11.5(4), 6.2(2))\,\mathrm{Hz}$. The s-wave scattering length is initially set to $a_s\!=\!300a_0$. To create an ensemble of independent 1D tubes, we adiabatically ramp up two horizontally oriented optical lattices along the $y$ and $z$ axes to a depth of $25\,\Er$ within $1\,\mathrm{s}$. At this depth, inter-tube tunneling is negligible, ensuring the realization of an array of vertically oriented 1D systems with an average occupancy of approximately $23$ atoms per tube. Due to the confinement by the lattice beams, the trapping frequency along the longitudinal $x$ direction increases to $\omega_x\!=\!2\pi \times14.1(4)\mathrm{Hz}$. The system is then driven into a 3D Mott insulating state by ramping a longitudinal lattice (i.e. along the $x$-axis) to $20\,\Er$. Subsequently, we tune the magnetic field to reach the target value for the scattering length $a_s$ via a Feshbach resonance. The effective 1D coupling constant $g_{\mathrm{1D}}$ is given by~\cite{olshanii1998}
\begin{equation}
g_\mathrm{1D} = \frac{2\hbar^2 a_{s}}{m a_\perp^2} \left( 1 - C \frac{a_{s}}{a_\perp} \right)^{-1}\text{,}
\end{equation}
where $m$ is the mass of the particles, $a_\perp \!=\! \sqrt{\hbar / m\omega_\perp}\!\approx\! 76\ \mathrm{nm}$ is the transverse harmonic oscillator length, and $C \!\approx\! 1.0326$ is a constant. The quench dynamics are initiated from a pre-equilibrated state at an $x$-lattice depth of $15\,\Er$, reached via a $200$-ms adiabatic ramp. The quench is then performed by abruptly lowering the $x$-lattice depth to $5\,\Er$ within $\SI{100}{\micro\second}$. After a variable hold time after the quench, we turn off all the optical fields and the interaction to perform a time-of-flight measurement with an expansion time of $50\,\mathrm{ms}$. The resulting density profile along the $x$-axis, which represents a sum over the contributions from all tubes, provides the momentum distribution $n(k, t)$. By scanning the hold time $t$, we obtain $n(k,t)$. The temporal evolution is sampled with a step size $\Delta t$ and a total duration $T$ chosen to resolve the characteristic energy scales of the post-quench Hamiltonian, which are typically determined by the energy gap.
%\todo{Introduce $g_{1D}$}

%%%%%%%%%%%%%%%%%%%%%%%%%
\subsection{Numerical methods and parameters}
\subsubsection{QMC calculations}
We calculate the energy gap for comparing to the measured dispersion via a path integral quantum Monte Carlo simulation, as shown in Fig.~\ref{fig:fig2}. We conduct simulations of the 1D bosonic Hamiltonian
\begin{equation}
\hat{H}\!=\!\sum_{i\!=\!1}^{N}\left[-\frac{\hbar^{2}}{2m}\nabla_{i}^{2}+V_{x}\sin^{2}\left(\frac{\pi x_{i}}{a}\right)\right]
+g_{1\mathrm{D}}\sum_{i<j}\delta(x_{i}-x_{j})
\label{eq1.Hamiltonian}
\end{equation} 
at a near-zero temperature of $T\!=\!0.5\,\mathrm{nK}$ with unit filling and chemical potential $\mu$ in a grand-canonical ensemble. The 1D coupling constant $g_{\mathrm{1D}}$ is given values that match the experimental condition. The interaction propagator is evaluated using
the pair-product approximation, which is exact for delta-function interactions~\cite{yan2015incorporating}. The MI phase is identified by the vanishing of the number compressibility, $\chi \!=\! \partial N/\partial \mu \!\to\! 0$. Consequently, the energy gap is extracted from the width of the incompressibility plateau in the $N(\mu)$ curve, defined by $\Delta \!=\! \mu_{\max} - \mu_{\min}$. Our numerical calculations are supported by the Algorithms and Libraries for Physics Simulations (ALPS) scheduler library and statistical analysis
tools~\cite{troyer1998parallel,albuquerque2007alps,bauer2011alps}.
\subsubsection{DMRG simulations}
In our DMRG simulations, we consider a one-dimensional Bose-Hubbard (BH) model (equation~(\ref{eq.1})) with $L\!=\!35$ sites in the presence of an external harmonic potential that matches the experimental conditions. The system is initialized at unit filling ($\bar n\!=\!1$), with the harmonic confinement identical to that for the experiment. The initial state is prepared as the ground state at a lattice depth of $V_z \!=\! 15\,\Er$. The corresponding BH parameters are obtained from numerical calculations of the lowest-band Wannier functions $\ket {w_{s,i}}$, yielding
\begin{align}
    U&=g_{1D}\int {\rm d}x |w_{s,i}(x)|^4 \label{eq.7}\\
    J&=-\langle w_{s,i}|\hat H_0|w_{s,i+1}\rangle \label{eq.8},
\end{align}
where $\hat H_0$ is the one-body Hamiltonian in the lattice.

The subsequent quench dynamics are simulated by solving the time-dependent Schrödinger equation associated with the post-quench BH Hamiltonian using the time-evolving block decimation (TEBD) method within the matrix product state (MPS) framework, as implemented in the TenPy~\cite{tenpy2024}.
On the SF side, the post-quench BH parameters $U$ and $J$ are also determined from equations~(\ref{eq.7}) and (\ref{eq.8}). On the MI side, we instead extract effective parameters from fits to the experimentally measured excitation spectra. Specifically, from the upper and lower edges of the dispersion, $\nu_{\max}$ and $\nu_{\min}$, we obtain
\begin{align}
    U/h &= \frac{\nu_{\max}+\nu_{\min}}2\label{eq.9}\\
    J/h &= \frac{\nu_{\max}-\nu_{\min}}{12}\label{eq.10}
\end{align}
The resulting QSF reproductions and the dynamics of $G^{(1)}(x,t)$ are shown in Figs.~\ref{fig:fig2} and \ref{fig:fig3}, respectively. For both the initial state preparation and the time evolution, we employ a truncation with maximum bond dimension $\chi_{\max}\!=\!500$ and local occupation cutoff $N_{\max}\!=\!5$, which ensures convergence of all reported results.

%%%%%%%%%%%%%%%%%%%%%%%%%
\subsection{Extraction of the sound velocity}

\begin{figure}
    \centering
    \includegraphics[width=0.85\linewidth]{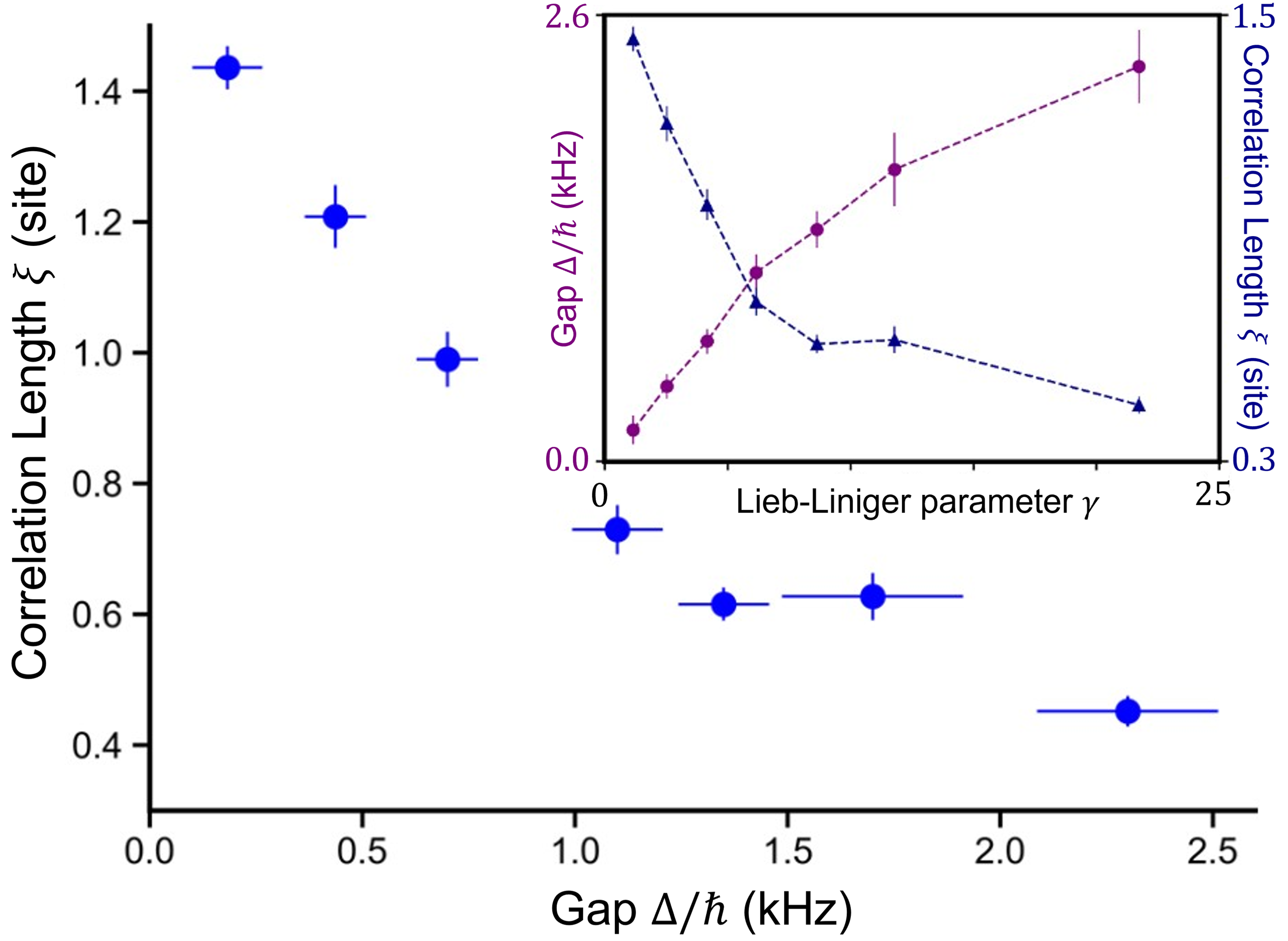}
    \caption{
    \textbf{Relations between gap, correlation length and interaction strength.} The blue dots illustrate the relationship between correlation length and gap. The error bars reflect the fitting uncertainties. Purple dots and dark blue triangles in the inset describe the relations between the gap with interaction and the correlation length with interaction, respectively. Dashed lines are drawn to guide the eye by connecting adjacent data points.} 
    \label{fig:figS1}
\end{figure}

In the Mott regime, the sound velocity is obtained from the relation between the excitation gap and the correlation length. 
Experimentally, we prepare the system at different values for the interaction strength $\gamma$ and measure the corresponding equilibrium momentum distributions $n(k)$. By Fourier transforming $n(k)$, we obtain the static one-body correlation function $G^{(1)}(x)$, which exhibits an exponential decay $G^{(1)}(x)\sim e^{-x/\xi}$. The correlation length $\xi$ is extracted by performing a linear fit to $\ln G^{(1)}(x)$ with the intercept fixed to zero.

Independently, the excitation gap is measured using LMS. The relation between the gap and the correlation length is shown in Fig.~\ref{fig:figS1}. As the interaction strength increases, the gap increases while the correlation length decreases; their dependence on $\gamma$ is summarized in the inset of Fig.~\ref{fig:figS1}.
Combining these independently measured quantities, we determine the sound velocity through the relation $1/\xi \!\sim\! \Delta/(\hbar v_s)$~\cite{Cazalilla2011RMP} for varying values of the interaction strength.

In the SF regime, the sound velocity corresponds to the slope of a linear dispersion in the Bogoliubov mode, which is predicted to be visible in quench spectrum by numerics. However, the observed spectrum is scarred by a few horizontal lines with an energy spacing roughly equal to the chemical potential. We attribute this effect to inter-tube interference, which is not captured in the 1D numerical simulation. The coherence between tubes enables inter-tube hopping, which produces energy offsets from the initial energy $E_0$ for certain tubes by approximately integer multiples of the chemical potential, i.e., $E \!\approx\! E_0 + n\mu$ with $n \in \mathbb{Z}$ representing the particle number acquired or lost, yielding spectral signals at energies $\Delta E\!\approx\!n\mu$.

%We attribute the most likely origin for this effect to the inter-tube hopping, which leads to energy shifts of integer time of chemical potential for certain tubes. The interference between tubes with and without hopping creates the fringes. }
%\todo{V1:We attribute this effect to weak inter‑tube hopping, which is not included in our one‑dimensional simulations. Such tunneling produces energy offsets in a subset of tubes by approximately integer multiples of the chemical potential, i.e., $E \approx E_0 + n\mu$ with $n \in \mathbb{Z}$. Interference between tubes that experience these offsets and those that do not yields the $\mu$-spaced energy fringes observed in the spectrum.

%%%%%%%%%%%%%%%%%%%%%%%%%
\subsection{Renormalization to s-band BHM}

\begin{figure}
    \centering
    \includegraphics[width=0.85\linewidth]{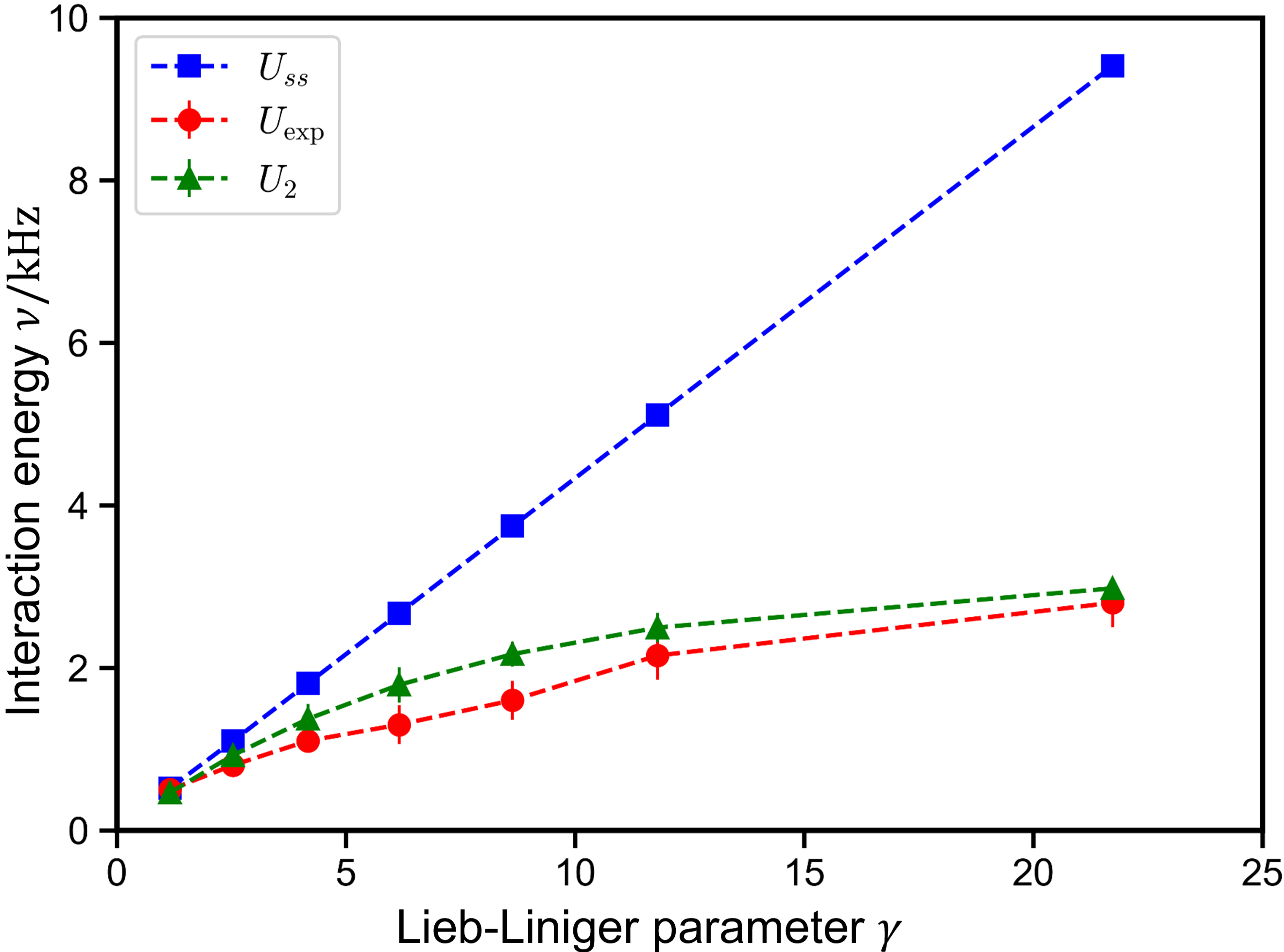}
    \caption{
    \textbf{Renormalization of the BH parameter $U$.} The red dots plots the experimentally extracted $U_{\rm exp}$. The blue squares and green triangles show the bare $s$-band parameter $U_{ss}$ and the effective two-body interaction $U_2$, respectively. Dashed lines are drawn to guide the eye by connecting adjacent data points.} 
    \label{fig:figS2}
\end{figure}

In the Mott regime, we extract the effective single-band BH interaction $U_{\rm exp}$ from equation (\ref{eq.9}). We next compare this experimentally determined parameter to the bare $s$-band values $U_{ss}$ computed from equation (\ref{eq.7}), as shown in Fig.~\ref{fig:figS2}. As the interaction becomes stronger, $U_{\rm exp}$ shows a large discrepancy with $U_{ss}$, indicating a breakdown of the bare s-band description as the interaction strength grows.\\

To clarify the origin of this deviation, we solve the exact two-body problem within a single lattice site, including multi-band effects. Defining the one-particle ground-state energy as $E_1$ and the two-particle ground-state energy as $E_2$, an effective two-body interaction is given by $U_2 \!=\! E_2 - 2E_1$. The two-body Schr\"odinger equation
\begin{align}
    &E\Psi(x_1,x_2) = \hat H \Psi(x_1,x_2)\nonumber \\
    =& \left[\sum_{i=1,2}\left(-\frac{\hbar^2\nabla_i^2}{2m}+V(x_i)\right)
    + g_{\rm 1D}\delta(x_1-x_2)\right]\Psi(x_1,x_2),
\end{align}
where the lattice potential, truncated in one site, $V(x)\!=\!V_x\sin^2{(k_Lx)}+V_0\theta(|x|-{\pi}/{k_L})$ with $\theta$ Heaviside step function and $V_0$ much larger than any energy scale in the system, is diagonalized on a truncated Wannier basis (up to $n_{\rm orb}=20$). 
The resulting $U_2$, as shown in Fig.~\ref{fig:figS2}, closely follows $U_{\rm exp}$ for $V_x\!=\!5\,\Er$, capturing the same interaction dependence. This agreement indicates that multi-band renormalization, already at the two-body level, plays a dominant role under our experimental conditions.

Within this framework, the experimentally extracted parameters $U$ and $J$ should be understood as renormalized BH parameters that incorporate higher-band dressing. Interaction-induced admixture of excited bands broadens the single-particle wave functions, effectively reducing the on-site interaction energy. Through this renormalization, the underlying multi-band model can be mapped onto a dressed single-band BH model that faithfully captures the low-energy excitation spectrum and dynamical transport properties of the system. Furthermore, this provides evidence that supports our use of bare s-band BH parameters in the simulation on the SF side in Fig.~\ref{fig:fig2}a. When the interaction is weak, these three results match quantitatively, indicating that the bare s-band description holds in this regime.

%%%%%%%%%%%%%%%%%%%%%%%%%
\subsection{Integrated one-body correlation and QSF}
Theory predicts that correlators after a global quantum quench in a translation-invariant system take the generic form \cite{Cevolani2015,Cevolani2016}
\begin{align}
    G(R,t)=g(R)-\int_{\rm BZ} \frac{{\rm d}{k}}{2\pi} \mathcal F_g ({k}) \frac{e^{i{kR}+2E_{ k}^ft}+e^{i{kR}-2E_{k}^ft}}{2},
\end{align}
where $2E_{k}^f$ is the dispersion of a pair of two excitations with opposite momenta of the final state, and the structure factor $\mathcal F_g ( k)$ denotes the contributions from the specific observable and the initial pre-quench state. In one dimension, the stationary phase condition is given by the solution of $\partial_k(kR\pm 2E_k^ft)\!=\!0$, which defines the group velocity as $v_g(k) \!=\! \partial_kE_k^f$. Assuming $v_g(k)$ has an upper bound $v_g^*$ at $k\!=\!k^*$, for $t<R/(2v_g^*)$ there is no stationary phase thus the correlation decays, indicating the envelope velocity is twice the maximum group velocity $2v_g^*$. In contrast, for $t\!>\!R/(2v_g^*)$ the stationary phase condition can be reached, and the correlation after subtracting $g(R)$ can be simplified under the  saddle-point approximation~\cite{villa_thesis} to
\begin{align}
    \Delta G(R,t)\propto \frac{\mathcal F_g (k^*)}{\left(|\partial_k^2 E_{k^*}|t\right)^{1/2}}\cos{\left( k^* R-2E_{k^*}^ft +\frac{\pi}{4} \right)},
\end{align}
which creates a sine-like signal with a phase-front propagating velocity $2v_p\!=\!2E_{k^*}/k^*$.
In the experiments, what we can get from the Fourier transformation of the momentum distribution is the integrated one-body correlation,
\begin{align}
    G^{(1)}(x,t)= \mathcal F_k \{n(k,t)\}\propto \int \langle\hat\Psi^\dagger(x+x')\hat\Psi(x') \rangle {\rm d} x'
\end{align}
which also encodes this two-fold structure in the correlation spreading when the external inhomogeneous potential is weak.

On the other hand, the same analysis along $t$ yields the QSF \cite{villa2019,yu2025},
\begin{align}
    S(k,\omega)=\mathcal F_t \{n(k,t)\}
=\sum_{n,n'}\mathcal{F}_{s}^{nn'}(k) \delta(E_n-E_{n'}-\hbar\omega)
\end{align}
where $E_n(k)$ is the dispersion of excitation pairs, and $\mathcal F_s^{nn'}(k)$ is another structure factor affected by the initial state and the finite temperature. The first order term of QSF directly encodes the energy spectrum $\delta(E_n(k)-\hbar \omega)$, providing an access to the full dispersion of quasiparticles.

%%%%%%%%%%%%%%%%%%%%%%%%%
\subsection{Extraction of the group velocity}
To directly extract the group velocity from the envelope of the propagating $G^{(1)}(x,t)$ signal, we preprocess the raw data to obtain a well-defined propagation front. 
In practice, we first smooth the measured correlation function $G^{(1)}(x,t)$ by applying a two-dimensional Gaussian convolution,
\begin{align}
G_{\mathrm{smooth}}(x,t)=\iint dx' dt'G^{(1)}(x',t')
\mathcal{G}_{\sigma_x,\sigma_t}(x-x',t-t'),
\end{align}
where the 2D Gaussian kernel is given by
\begin{align}
\mathcal{G}_{\sigma_x,\sigma_t}(x,t)=
\frac{1}{2\pi\sigma_x\sigma_t}
\exp\!\left(-\frac{x^2}{2\sigma_x^2}-\frac{t^2}{2\sigma_t^2}\right).
\end{align}

The smoothing widths $\sigma_x$ and $\sigma_t$ are chosen to match the spatial and temporal resolution of the data (corresponding to two pixels in each direction).
To isolate the envelope, we then apply a thresholding procedure by setting all values below a small cutoff $\varepsilon$ to zero to suppress low-amplitude noise,
\begin{align}
G_{\mathrm{proc}}(x,t)=
\begin{cases}
G_{\mathrm{smooth}}(x,t), & |G_{\mathrm{smooth}}(x,t)|\ge \varepsilon,\\
0, & |G_{\mathrm{smooth}}(x,t)|<\varepsilon.
\end{cases}
\end{align}

The cutoff $\varepsilon$ is chosen within the same order of magnitude for all interaction strengths, with slight adjustments depending on the signal amplitude.
The resulting processed signal $G_{\mathrm{proc}}(x,t)$ exhibits a well-defined envelope boundary, from which the group velocity is extracted by tracking the propagation of the leading edge of the envelope, defined by the outermost nonzero contour of $G_{\mathrm{proc}}(x,t)$.
We find that the extracted group velocity shows a moderate dependence on the precise choice of $\varepsilon$; this sensitivity is taken into account as a systematic contribution to the uncertainty, resulting in an estimated $\sim15\%$ error bar for $v_g$.

%%%%%%%%%%%%%%%%%%%%%%%%%
\subsection{Velocities in the deep MI regime}
Deep in the Mott regime with $\bar n\!=\!1$, the excitation spectrum may be written in terms of particle-hole pairs with opposite momenta. It yields the massive relativistic particle-like dispersion relation~\cite{Peter2012}
\begin{align}
    2E_{ph}(k)
    = \sqrt{(U-6J)^2 + 24 J U \sin^2(k/2) - 4 J^2\sin^2(k)}.
\end{align}
In the low-$k$ limit, it reduces to the bare relativistic formula $2E_{ph}(k) \!\approx \!\sqrt{\Delta^2+ (\hbar v_s k)^2}$,
where the gap $\Delta\!\approx\!U-6J$ plays the role of the mass energy and $v_s\!\approx\!\sqrt{6JU}/\hbar$ that of the speed of light. In the context of correlated quantum matter, $v_s$ is rather referred to as the speed of sound since, at the Mott critical point where $\Delta\!=\!0$, it yields the slope of the phonic branch $2E_{ph}(k) \approx \hbar v_s k$.

In the limit $U \!\gg\! J$, the (maximum) group velocity $v_g \!=\! \partial 2E_{ph}(k) / \partial \hbar k$ is found at $k^* \!=\! \pi/2$ and reads as
\begin{align}
    v_g^* = 6J/\hbar+\mathcal O(\frac{J^3}{\hbar U^2}).
\end{align}
The corresponding phase velocity $v_g \!=\! 2E_{ph}(k) / \hbar k^*$ is then
\begin{align}
    v_p = \frac{2 U}{\pi\hbar}.
\end{align}
From these two expressions, we obtain the relativistic-like invariance relation
\begin{align}
    v_g v_p =\frac{2 v_s^2}{\pi}.
\end{align}
This relation is valid at dominant order in $J/U$.

%\sout{In the deep Mott regime with $\bar n\!=\!1$, the dispersion relation can be simplified to}
%\begin{align}
%    E_{ph}(k)
%    =U-6J\cos k+\frac{7}{8}\frac{J^2}{U}\cos^2 k+\mathcal O(\frac{J^3}{U^2})
%\end{align}

%\sout{The sound velocity is defined as the coefficient in the relativistic-like dispersion when $k\to0$. By expanding $E_{ph}(k)$ around $k\!=\!0$ we obtain}
%\begin{align}
%    2v_s=\sqrt{6UJ}-\sqrt{\frac{2J}{3U}}J+\mathcal{O}(\frac{J^{\frac{5}2}}{U^{\frac{3}2}})
%\end{align}

%\sout{The (maximum) group velocity is defined as that with characteristic momentum $k^*$,}
%\begin{align}
%    2v_g^*=2v_g(k^*)=6J+\mathcal O(\frac{J^3}{U^2})
%\end{align}

%\sout{And the phase velocity is defined at the characteristic momentum $k^*$, which reads:}
%\begin{align}
%    2v_p=\frac{E_{ph}(k^*)}{k^*}=\frac{2}{\pi}U-\frac{7}{6\pi^2}J+\mathcal O(\frac{J^2}{U})
%\end{align}

%\sout{In the main text, we denote those three velocities under the neutral excitation representation, which means $\{2v_p,2v_s,2v_g^*\} \to \{v_p,v_s,v_g\}$.
%From these expressions, we can obtain the relativistic-like invariance:}
%\begin{align}
%    v_g=\frac{2}\pi\frac{v_s^2}{v_p}\pm\Delta,
%\end{align}
%\sout{where the remainder satisfies}
%\begin{align}
%    |\Delta|\leq\frac{7}{9\pi^4}\left(\frac{v_s}{v_p}\right)^4v_p,
%\end{align}
%\sout{which means that the discrepancy to the relativistic-like invariance is}
%\begin{align}
%    |\frac{v_s^2}{v_pv_g} - \frac{\pi}{2}|=\frac{\pi}{2}\frac{|\Delta|}{v_g}\le \frac{7}{18\pi^3}\frac{v_s^4}{v_p^3v_g}\lesssim \frac{7}{24}\frac{J}{U}
%\end{align}

%%%%%%%%%%%%%%%%%%%%%%%%%
\end{document}